\documentclass[proof]{WileyASNA-v1}
\usepackage{color}

\articletype{Comment}%

\received{20 May 2020}
\revised{17 August 2020}
\accepted{}

\begin{document}

\title{A resolution of the Trans-Planckian problem in the $R_{\rm h}=ct$ universe}

\author[1]{Fulvio Melia*}

\authormark{Fulvio Melia}

\address[1]{\orgdiv{Departments of Physics and Astronomy, and the Applied Math Program}, 
\orgname{The University of Arizona}, \orgaddress{\state{Tucson, Arizona}, \country{U.S.A.}}}

\corres{*\email{fmelia@arizona.edu}}


\abstract{The recent measurement of a cutoff $k_{\rm min}$ in the
fluctuation power spectrum $P(k)$ of the cosmic microwave background
may vitiate the possibility that slow-roll inflation can simultaneously
solve the horizon problem and account for the formation of structure via
the growth of quantum fluctuations in the inflaton field. Instead, we
show that $k_{\rm min}$ may be interpreted more successfully in the
$R_{\rm h}=ct$ cosmology, as the first mode exiting from the Planck scale
into the semi-classical Universe shortly after the Big Bang.
In so doing, we demonstrate that such a scenario completely
avoids the well-known trans-Planckian problem plaguing standard
inflationary cosmology.} 

\keywords{cosmological parameters; cosmology: observations; cosmology: early Universe;
cosmology: theory; gravitation; cosmology: inflation}

\maketitle

\section{Introduction}
The Friedmann-Lema\^itre-Robertson-Walker (FLRW) metric, based on
the cosmological principle and its assumption of isotropy and homogeneity
on large scales, is the backbone of modern cosmology. All the available
observational evidence appears to support its essential spacetime basis
\citep{Melia:2020}, so any conceptual or foundational hurdles arising with 
the expansion of the Universe are attributed to other factors---notably
an incomplete understanding of the physics underlying the evolution of
its contents.

Over the past four decades, several crucial amendments and additions have
been introduced to the basic picture in order to address some of these difficulties,
chief among them the well-known horizon problem associated with the uniformity
of the cosmic microwave background (CMB) temperature, $T_{\rm cmb}$. In the
context of $\Lambda$CDM, CMB photons emitted near the surface of last scattering
(LSS) at redshift $z\sim 1080$ from opposite sides of the sky would be causally
disconnected without an anomalous accelerated expansion in the early Universe \citep{Melia:2013a}.
Yet $T_{\rm cmb}$ has the same value in all directions, save for $\sim 10^{-5}$
variations associated with fluctuations seeded at, or shortly after, the Big Bang.

A very elegant solution to this problem was introduced in the early 1980's
\citep{Guth:1981}, based on an expected phase transition in grand unified theories
(GUTs), when the strong and electroweak forces may have separated at an energy scale
$\sim 10^{16}$ GeV, or $\sim 10^{-35}$ seconds after the Big Bang. As long
as the scalar field, $\phi$, associated with this spontaneous symmetry breaking
had the `right' potential, $V(\phi)$, one could envisage an evolution at almost
constant energy density, $\rho_\phi$, producing a transient near-de Sitter cosmic
expansion \citep{Linde:1982}. Such an inflationary phase would have exponentially
stretched all observable features well beyond the Hubble radius, $R_{\rm h}=c/H$,
where $H(z)$ is the redshift-dependent Hubble parameter, causally connecting the
spacetime throughout the visible Universe today.

Perhaps even more importantly, this event is believed to have also produced
the large-scale structure via the seeding of quantum fluctuations in $\phi$
and their subsequent growth to classically relevant scales during the inflated
expansion \citep{Kodama:1984,Mukhanov:1992}. A near scale-free 
spectrum $P(k)$ would have been generated as modes with comoving wavenumber 
$k$ successively crossed $R_{\rm h}$ and classicalized, freezing their amplitude 
at a mode-dependent crossing time $t_k$ \citep{Maldacena:2003}. Thus, 
inflation appears to have simultaneously solved the $T_{\rm cmb}$
horizon problem and provided an explanation for the origin of $P(k)$.

In spite of this initial success, however, the inflationary paradigm is
nonetheless conceptually incomplete for several reasons. For example, the
recent discovery of the Higgs particle \citep{Aad:2012} has reminded us that
$\Lambda$CDM is subject to several horizon problems---at several different
epochs---not just one, so the GUT transition at $\sim 10^{-35}$ seconds
is looking more like an overly customized solution focusing solely on $T_{\rm cmb}$,
rather than providing an over-arching paradigm to account for our entire
causally-connected Universe. A second well-motivated phase transition should have
occurred when the electric and weak forces separated at a critical temperature
$T_{\rm Higgs}\sim 159.5\pm 1.5$ GeV, i.e., $t\sim 10^{-11}$ seconds after the
Big Bang \citep{Fileviez:2009,Weir:2018,Noble:2008,Dolan:2012,Barr:2015}, too
far beyond the GUT scale to have been affected by the hypothesized first transition
\citep{Melia:2018a}. This second spontaneous symmetry breaking would have
inevitably led to its own horizon problem, having to do with the `turning on'
of the Higgs field and its vacuum expectation value, which today appears to
be universal, even on scales exceeding the regions that were causally
connected at the time of the electroweak phase transition.

This is not so much an argument against inflation per se, though it does weaken
the claim that a GUT phase transition could account for all of the major features
we see today; it apparently does not. A more serious problem with the slow-roll
inflationary paradigm has been uncovered by another recent study of the angular
correlation function measured in the CMB by {\it Planck} \citep{Planck:2018}.

The inflationary model faces several other hurdles, not merely possible
inconsistencies with the horizon problems. For example, recent observations of the
CMB anisotropies have apparently eliminated a broad range of possible inflationary
scenarios, disfavoring all the simplest potentials \citep{Ijjas:2013}. And the remaining 
range of inflaton fields are subject to serious drawbacks, making it much more difficult 
to understand how the universe acquired its initial conditions. Along with 
this outcome, many of the best-motivated inflationary scenarios appear to have been
ruled out, as we shall see below. According to this recent analysis 
\citep{MeliaLopez:2018,LiuMelia:2020}, none of the slow-roll potentials based on
the `Quadratic,' `Higgs-like' and `Natural' forms can account for the observed 
power spectrum while simultaneously mitigating the horizon problem.
Having said this, it is not really our goal in this {\it Note} to argue
against the inflationary paradigm, though the difficulties faced by inflationary
cosmology certainly add to our motivation for pursuing the alternative interpretation 
described here.

In the standard picture, the solution to the $T_{\rm cmb}$ horizon problem, and a 
generation of a near scale-free fluctuation spectrum $P(k)$, are intimately connected 
via the initiation and extent of the inflationary phase. But the CMB
angular-correlation function now provides compelling evidence---at a
confidence level {\it exceeding} $8\sigma$---that $P(k)$ has a non-zero
cutoff $k_{\rm min}={(4.34\pm0.50)/r_{\rm cmb}}$, where $r_{\rm cmb}$
is the comoving distance to the LSS \citep{MeliaLopez:2018}. Since
$k_{\rm min}$ would have been the first mode crossing $R_{\rm h}$
during inflation, it would signal the precise time, as a function of $V(\phi)$,
at which the de Sitter expansion started. Unfortunately, its measured
value shows that none of the slow-roll potentials proposed thus far can
simultaneously account for the uniformity of $T_{\rm cmb}$ across the sky
and the observed $P(k)$ in the CMB \citep{LiuMelia:2020}. The conclusion
from this is that, if slow-roll inflation is to work, it must function
in a more complicated way than has been imagined thus far.

As we shall see in this {\it Note}, the measurement of $k_{\rm min}$ in
the angular correlation function of the CMB not only constrains the time
when inflation could have started, but apparently provides direct evidence
of quantum fluctuations at the Planck scale. This topic broaches one of
the most serious fundamental problems with inflation, one that
has eluded satisfactory resolution for over three decades. It is generally
understood that to solve the horizon problem in $\Lambda$CDM, a minimum of
60 e-folds of inflationary expansion must have occurred, even more in
many variants of the basic model. Thus, cosmological scales of observational
relevance today must have expanded from sub-Planckian wavelengths at the
start of inflation \citep{Brandenberger:2001,Niemeyer:2001,Kempf:2001,Easther:2001}.
But the physics we have today cannot adequately handle such processes,
a situation known as the `trans-Planckian problem' (TP) \citep{Martin:2001}.
(See \cite{Brandenberger:2013} for a detailed review.) This 
signals a potentially fatal incompleteness of inflationary theory at a
fundamental physics level.

The key here is how one should interpret the evolution of
quantum fluctuations seeded well below the Planck scale, where quantum mechanics
and general relativity as we know them today are probably not valid
\citep{NiemeyerRenaud:2001}. The problems that derive from attempting to
follow this trans-Planckian evolution have been well documented. For example,
attempts to renormalize a scalar theory with nontrivial initial conditions
in flat space run into possible divergences confined to the surface where
the initial conditions are imposed \citep{Collins:2005}. 

As we shall see
in the next section, the development of a quantum theory of gravity could
resolve such issues, allowing us to follow the growth and evolution of 
quantum fluctuations from birth to their exit into the semi-classical
universe (beyond the Planck domain). Attempts have also been made to modify
the dispersion relation for quantum modes on short scales (see, e.g., 
refs.~\citealt{Ashoorioon:2011,Zhu:2016}), or to alter the Heisenberg uncertainty
relation \citep{Easther:2001}, in order to uncover possible observational 
signatures that one may detect in the CMB, allowing us to make some
progress in identifying new physics at the Planck scale.

Our goal in this {\it Note} is to provide an alternative interpretation
of the cutoff $k_{\rm min}$ measured in the primordial power spectrum
$P(k)$, in order to mitigate this current need of having to follow the
evolution of inflationary quantum fluctuations where our semi-classical 
theories are unlikely to be valid. 

\section{Phenomenological Approach to TP Physics in $\Lambda$CDM}
One can easily understand why our inability to analyze physics below the 
Planck scale constitutes a potentially insurmountable problem. The
Planck mass is a unit of mass defined solely using fundamental and universal units,
and is given by $M_{\rm P}\equiv \sqrt{\hbar c/G}$. As it turns out, the Planck
length associated with $M_{\rm P}$ is of the same order as the Schwarzshild radius
and the Compton wavelength of the Planck mass, suggesting that it represents a
transition from classical general relativity to the quantum gravity domain. Given
this `physical' interpretation, we shall use in this {\it Note} a slightly different 
value of the Planck mass, which we shall call $m_{\rm P}=\sqrt{\pi\hbar c/G}$ (see
below), derived from strictly setting its Compton wavelength equal to its Schwarzschild 
radius.

It is not difficult to see that, since the former length scale increases as the 
latter shrinks towards the Big Bang, it is simply not possible to characterize 
the behaviour of modes below the Planck scale using quantum mechanics and general 
relativity separately. The semi-classical physics we use to describe the evolution 
of quantum fluctuations as the Universe expands does not apply for mode scales
shorter than their Compton wavelength \citep{Brandenberger:2013}.

This problem manifests itself in several ways, particularly via the
mode normalization that one must use to calculate $P(k)$ for a
comparison with the CMB data. The amplitude of the modes is typically
inferred by minimizing the expectation value of the Hamiltonian, but
with a time-dependent spacetime curvature at the Planck scale, the
frequencies themselves depend on time and non-inertial effects. Early
attempts at addressing this issue extended the birth of fluctuation modes
into the very distant conformal past, well below the Planck scale,
arguing that the simple harmonic oscillator is recovered there, allowing
one to impose a Minkowski vacuum---called the Bunch-Davies vacuum in this
context \citep{Bunch:1978}---as the background for the fluctuations. But given
that the physics below the Planck scale is unknown, we have a conceptual
problem understanding whether or not the Bunch-Davies vacuum is even
the correct choice for sub-Planckian modes.

We should point out, however, that this trans-Planckian problem
arises primarily from the use of quantum mechanics with (the classical theory
of) general relativity. It is believed that this problem may disappear with a 
viable theory of quantum gravity, such as Wheeler-DeWitt theory, loop quantum gravity
or string theory \citep{Kiefer:2004}. Significant effort is being expended in
studying such models, which are established by using quantization techniques 
on symmetry-reduced general relativity. For a large class of wave functions,
the homogeneous and isotropic FLRW metric even loses its singularity
\citep{Ashtekar:2008}. 

But such an approach is not entirely problem free either. These models suffer 
from the long-standing measurement problem in quantum mechanics, 
having to do with the ambiguity of when exactly collapses happen and how a unique,
determanistic outcome is selected. The measurement problem is even more severe in 
cosmology, because the universe lacks outside observers or measurement devices 
that could have collapsed the wave function. There is also the problem of time,
given that the wave function is static in both the Wheeler-DeWitt theory and 
loop quantum gravity \citep{Kiefer:2004,Kuchar:2011}, making it difficult to envisage
temporal evolution, e.g., to ascertain whether the universe is expanding or contracting.
Finally, such quantum gravity theories are based solely on a wave function, without
an actual metric, so it is difficult to determine the behavior of quantum
fluctuations in the remote conformal past, where singularities might have emerged 
anyway \citep{DeWitt:1967}.

The consensus today is that Planck-scale physics probably should have created
an imprint on the CMB, but with no established theory of quantum gravity,
no one knows how to predict such features with any confidence. Instead,
the approaches taken over the past two decades have been based
on phenomenological treatments, including (1) modifications to the
dispersion relation for quantum modes on short scales
\citep{Brandenberger:2001,Martin:2001,Niemeyer:2001,Joras:2009,Ashoorioon:2011,Zhu:2016};
(2) the use of string-inspired changes to the Heisenberg uncertainty
relation \citep{Kempf:2001,Easther:2001,Hassan:2003}; and (3) noncommutative
geometry \citep{Chu:2001,Lizzi:2002,Brandenberger:2002}.

All of these are really probes of the CMB to suggest how basic theory ought to
be modified rather than robust attempts at using a well-justified model of physics
at short distances to predict a trans-Planckian signature. To understand the scale
we are considering here, we define the Planck length $\lambda_{\rm P}$ to
be the Compton wavelength $\lambda_{\rm C}\equiv 2\pi/m_{\rm P}$ of a (Planck)
mass $m_{\rm P}$ for which $\lambda_{\rm C}$ equals its Schwarzschild radius
$R_{\rm h}\equiv 2Gm_{\rm P}$. The Planck energy is therefore
$m_{\rm P}\approx 1.22\times 10^{19}$ GeV. Estimates of how big
trans-Planckian corrections might be, based on the above phenomenological
approaches, range from $(\lambda_{\rm P}/R_h[t_{\rm inf}])^2$ (see, e.g.,
\citealt{Kempf:2001b,Kaloper:2002}) to as large as $\lambda_{\rm P}/R_h(t_{\rm inf})$
\citep{Easther:2001,Easther:2002,Brandenberger:2002b,Danielsson:2002}. In these
expressions, $R_{\rm h}(t_{\rm inf})$ is the Hubble radius during inflation
(which is more or less constant in the slow-roll approximation).

Thus, if inflation is associated with a GUT phase transition at $\sim 10^{16}$ GeV,
these phenomenologically motivated corrections fall in the range $10^{-6}$ to $10^{-3}$.
Additional support for such a claim---especially towards the high-end of this range---is
provided by arguments \citep{Brandenberger:2002b,Danielsson:2002} that curvature
effects at the Planck scale probably produce deviations of the $\phi$ quantum state
from the local vacuum state on the order of $\lambda_{\rm P}/R_h(t_{\rm inf})$,
but no one really knows for sure. If reasonable, this range includes effects
potentially large enough to affect the primordial power spectrum $P(k)$ in
measurable ways (see, e.g., \citealt{Easther:2002}). Of course, a 
final resolution of whether or not trans-Planckian effects manifest themselves 
observationally must await the formulation of a well-motivated quantum gravity 
theory.

Certainly, the measurement of $k_{\rm min}$ already seems to argue 
against the premise that inflation might have started early enough to solve the 
temperature horizon problem, while simultaneously explaining the origin of $P(k)$. 
All three major satellite missions designed to study the CMB---COBE 
\citep{Hinshaw:1996}; WMAP \citep{Bennett:2003}; and {\it Planck} 
\citep{Planck:2018}---have uncovered several large-angle anomalies with the 
temperature fluctuations, including missing correlations at angles exceeding
$\sim 60^\circ$ and an unexpectedly low power in multipoles $2\lesssim \ell\lesssim 5$.
These unexpected features stand in sharp contrast to the general level of success
interpreting the CMB anisotropies with $\Lambda$CDM at angles $\lesssim 2^\circ$,
i.e. at $\ell\gtrsim 10$.

After a careful re-analysis of the latest release of the {\it Planck} data, it
now appears that both of these discrepancies have common origin: a sharp cutoff
of the primordical power spectrum $P(k)$ at the aforementioned minimum wavenumber
$k_{\rm min}$. In the first of these studies \citep{MeliaLopez:2018}, such a
cutoff was shown, not only to suppress the expected correlation at large angles but,
to actually significantly improve the fit of the angular correlation function
at all angles. This analysis demonstrated that a zero $k_{\rm min}$ was ruled out
at a confidence level exceeding $8\sigma$. The second study focused on the impact
of $k_{\rm min}$ on the CMB angular power spectrum itself, and concluded that
the missing power at low multipoles was equally well explained by this cutoff,
while none of the other 11 or 12 $\Lambda$CDM parameters optimized by {\it Planck}
were materially affected by the introduction of $k_{\rm min}$. 

The combined result of these two analyses is that a single value of $k_{\rm min}$
completely mitigates both large-angle anomalies. But as shown in \cite{LiuMelia:2020}, 
the interpretation of this cutoff as the first mode to cross the Hubble radius
$R_{\rm h}$ once slow-roll inflation began is then inconsistent with the
accelerated expansion required to provide us with a causally-connected Universe
today. The principal reason is that the time associated with this crossing was too
late to have permitted the quasi-de Sitter phase to have inflated the universe
sufficiently to solve the CMB temperature horizon problem. In the next section, 
we present an alternative interpretation of $k_{\rm min}$ that avoids these 
conceptual problems and, at the same time, completely eliminates the trans-Planckian 
inconsistency.

\section{A Resolution of the TP Problem in $R_{\rm h}=ct$}
The FLRW cosmology known as the $R_{\rm h}=ct$ universe
\citep{Melia:2007,Melia:2016a,Melia:2017a,Melia:2020,MeliaAbdelqader:2009,MeliaShevchuk:2012} 
is essentially $\Lambda$CDM, though with an additional constraint motivated by both the
observational evidence and a careful application of the Local Flatness Theorem in
general relativity. It too has an energy density $\rho$ dominated by various combinations
of matter, $\rho_{\rm m}$, radiation, $\rho_{\rm r}$ and dark energy, $\rho_{\rm de}$,
depending on the cosmic epoch, but the equation-of-state of this `cosmic fluid' always satisfies
the so-called zero active mass condition, $\rho+3p=0$, where $\rho=\rho_{\rm m}+\rho_{\rm r}
+\rho_{\rm de}$ and $p$ is the corresponding total pressure $p=p_{\rm m}+p_{\rm r}+p_{\rm de}$.
An introductory review of this model may be found in ref.~\cite{Melia:2019a}, and a more 
thorough presentation of its foundational support, both observational and theoretical, will 
be presented in the upcoming monograph \cite{Melia:2020}, to be released in Fall, 2020.

By now, this model has been subjected to comparative tests with basic $\Lambda$CDM 
using over 27 different kinds of cosmological data, and has accounted for the
observations at least as well as the standard model, actually better than latter in the
majority of cases. A recent summary of these comparative tests may be found in Table~2 of 
\cite{Melia:2018b}.

The $R_{\rm h}=ct$ cosmology was first postulated on the basis of such
empirical evidence showing that the {\it apparent} horizon $R_{\rm h}\equiv c/H$, averaged
over a Hubble time, equals the comoving distance, $ct$, light could have traveled since
the Big Bang \citep{Melia:2018b}. In general relativity, this horizon separates null
geodesics receding from the observer from those approaching him/her, but is
generally not the same as the event horizon, which signals the causal limit in the
asymptotic future \citep{Melia:2018c}. As such, there is no reason why $R_{\rm h}$
should always equal $ct$, unless there exists some specific, fundamental reason
forcing this condition.

Recently, the theoretical basis for $R_{\rm h}=ct$ was strengthened considerably
with a thorough re-examina\-tion of the lapse function (i.e., the coefficient
$g_{tt}$ in the FLRW metric; \citealt{Melia:2019b}. The issue of whether or not a
lapse function $g_{tt}=1$ is consistent with a non-inertial Hubble flow has been paid 
scant attention in the past. In this recent work, the Local Flatness Theorem 
\citep{Weinberg:1972} in general relativity was used to prove that $g_{tt}=1$ is in fact
valid for only two specific equations of state: an empty Universe with 
$\rho=p=0$ (i.e., Minkowski space) and the aforementioned zero active mass condition,
$\rho+3p=0$. As we now know, a pressure $p=-\rho/3$ produces a constant expansion
rate and, more importantly, forces the equality $R_{\rm h}=ct$. The empirical evidence
pointing to this constraint was therefore a pre-confirmation of the subsequent
theoretical analysis based on the Local FLatness Theorem, showing that 
the use of FLRW is valid only when the cosmic fluid satisfies this equation of state.

A notable feature of the expansion implied by this scenario is that it lacks
any horizon problem, eliminating the need for an inflated expansion of the early
Universe. Thus, if the zero active mass condition was evident at the earliest times,
$t$, it is straightforward to show \citep{Melia:2017b} that an incipient (though
non-inflationary) scalar field $\phi$ would have had the well-defined potential
\begin{equation}
V(\phi)=V_0\,\exp\left\{-{2\sqrt{4\pi}\over m_{\rm P}}\,\phi\right\}\;.
\end{equation}
$\phi$ is therefore a special member of the class of minimally coupled fields explored
in the 1980's, that produced power-law inflation \citep{Abbott:1984,Lucchin:1985,Barrow:1987,Liddle:1989}
except that, with the zero active mass equation-of-state, this $\phi$ produced a constant
expansion rate $a(t)=t/t_0$ and did not inflate.

In $R_{\rm h}=ct$, quantum fluctuations in $\phi$ with a wavelength $\lambda_k<2\pi R_{\rm h}$,
where $k$ is the comoving wavenumber and $R_{\rm h}$ is the Hubble radius, oscillate, while those
with $\lambda_k>2\pi R_{\rm h}$ do not \citep{Melia:2017b}. Thus, mode $k$ oscillated in the
semi-classical Universe once it emerged across the Planck scale. But the critical
question is ``When did it emerge?" From the expression $k=2\pi a(t)/\lambda_k(t)$, it is
clear that the observed value of $k_{\rm min}$ indicates the time $t_{\rm min}$ when the
first mode appeared. Therefore,
\begin{equation}
t_{\rm min}={4.34\,t_{\rm P}\over \ln(1+z_{\rm cmb})}\;,
\end{equation}
in terms of the redshift, $z_{\rm cmb}$, at the surface of last scattering.

In the concordance $\Lambda$CDM model, $z_{\rm cmb}\sim 1080$, for which
$t_{\rm min}\sim 0.6 t_{\rm P}$. With the expansion scenario implied by
$R_{\rm h}=ct$, this redshift could be quite different, but the dependence
of $t_{\rm min}$ on the location of the last scattering surface is so weak,
that even a redshift $z_{\rm cmb}\sim 50$ would result in an initial
emergence time of $t_{\rm min}\sim 1.1 t_{\rm P}$. Therefore, it appears that
$k_{\rm min}$ in $R_{\rm h}=ct$ represents the first mode exiting the Planck
region at about the Planck time, a compelling indication that the cutoff
$k_{\rm min}$ corresponds to the first mode that could have physically
emerged into the semi-classical Universe after the Big Bang.

Unlike the situation with an inflaton field, in which these modes were seeded in the
Bunch-Davies vacuum and oscillated across the trans-Planckian region, the quantum
fluctuations associated with a non-inflationary scalar field in the $R_{\rm h}=ct$
cosmology could well have been formed at the Planck scale and then evolved according
to standard physical principles in the semi-classical Universe. Such an idea---that
modes could have been created at a particular (perhaps even fixed) spatial
scale---is not new. It has been proposed and discussed in the literature before,
notably by \cite{Hollands:2002}, but also by \cite{Brandenberger:2002}
and \cite{Hassan:2003}, among others.

Whereas inflaton quantum fluctuations would have been born randomnly at
various times in the distant conformal past, our new proposal is that all of the non-inflaton 
fluctuations emerged into the semi-classical universe at the {\it same spatial scale},
i.e., the Planck length. They could very well have had a past history prior to reaching
$\lambda_{\rm P}$, but the difference is that we don't need to know how to dynamically
evolve them below this scale using our current version of quantum mechanics and general
relativity in order to self-consistently evolve them at $t\ge t_k$. In contrast, the
conventional picture requires that we use these semi-classical theories in a domain
where they are probably not valid. Without invoking a Bunch-Davies vacuum and following
their evolution prior to $t_{\rm P}$, we cannot ensure that the inflaton field simultaneously
fixed the horizon problem and produced the required primordial power spectrum $P(k)$.

Another key difference between the quantum fluctuations associated with an inflaton field
and those generated with the field potential in Equation~(1), is that the solution to the
Mukhanov-Sasaki equation \citep{Mukhanov:2005,Kodama:1984}, using an expansion factor
$a(t)=t/t_0$ consistent with this potential, can be easily shown to have a {\it constant}
frequency---dependent only on the time $t_k$ at which the quantum fluctuation emerged
out of the Planck domain. It was therefore a true harmonic oscillator and its amplitude
may be determined via canonical quantization in flat spacetime or, equivalently, by 
minimizing the Hamiltonian. This is due entirely to the zero active mass condition,
which produced zero acceleration, even at the Planck scale. Compare this with the 
conventional case, in which the extreme spacetime curvature at $\lambda_{\rm P}$ 
obviates such a straightforward approach and one must instead evolve the quantum
fluctuations starting with the Bunch-Davies vacuum in the very remote past.

Assuming that all subsequent modes continued to emerge across the Planck
scale with a wavelength $\lambda_k=2\pi\lambda_{\rm P}$, though at progressively
later times $t_k\equiv k\lambda_{\rm P}t_0$, it is trivial to show \citep{Melia:2017b}
that the resultant power spectrum is almost scale free, with an index $n_s$ slightly
less than one, consistent with the value measured by {\it Planck} \citep{Planck:2018}.
Thus, a non-inflationary scalar field in the $R_{\rm h}=ct$ universe can account
for both the measured cutoff $k_{\rm min}$ and for the observed distribution
of fluctuations in the CMB. Most importantly for the main theme of this paper,
the first reliable measurement of a minimum cutoff in the power spectrum $P(k)$
signals a direct link between the CMB anisotropies---and the subsequent formation
of structure in the Universe---and quantum fluctuations at the Planck scale. In
so doing, this interpretation eliminates one of the principal inconsistencies
with the basic slow-roll inflationary model, i.e., the well-known trans-Planckian
problem.

In short, this principal distinction
between the standard inflationary scenario and the mechanism described here is
that quantum fluctuations in the former had to be seeded on scales well below the
Planck length, at a time well before the Planck time in order to produce the
observed anisotropies in the CMB, while in the latter they could have formed at
the Planck scale, with the first emerging at the Planck time. All subsequent
modes would also have formed at the Planck scale, though after the Planck time,
thereby producing a near scale free spectrum, and always evolving dynamically
according to standard quantum mechanics and general relativity in the semi-classical
Universe. The mechanism we are describing here therefore completely avoids
the trans-Planckian problem, because our treatment of the quantum fluctuations
relies solely on the physics we know, based on initial conditions at the Planck
scale---not below it.

\section{Conclusion}
In this paper, we have discussed the implications of the fact that, in addition to
the well-studied power spectral index $n_s$ and amplitude of the CMB fluctuations,
we now have a robust measurement of a third parameter characterizing the primordial
perturbation spectrum, i.e., the wavenumber cutoff $k_{\rm min}$, which differs
from zero at a confidence level exceeding $8\sigma$ \citep{MeliaLopez:2018}.
This cutoff appears to invalidate basic slow-roll inflationary models attempting to 
simultaneously account for the 60 e-folds of exponential expansion at the GUT scale and
the generation of anisotropies in the CMB from quantum fluctuations in the
inflaton field \citep{LiuMelia:2020}. An additional well-known inconsistency 
with this scenario is the trans-Planckian problem, referring to the required transition of 
modes from below the Planck scale into the semi-classical Universe, a process that
cannot adequately be described by quantum mechanics and general relativity separately.

Contrasting with this deficiency in the standard model, we have also
demonstrated that the interpretation of $k_{\rm min}$ in the $R_{\rm h}=ct$
cosmology suggests it corresponds to the first quantum fluctuation that
could have physically emerged from the Planck scale shortly after the Big Bang.
This scenario thus avoids the trans-Planckian problem if one invokes the idea
that all fluctuations in the incipient scalar (though non-inflationary) field
were seeded at a fixed spatial scale---in this case, the Planck scale---though
at progressively later times depending on the wavenumber $k$ of the mode. This
interpretation is fully consistent with the quantum mechanical meaning of
the Planck length, representing the shortest physical size of any causally
connected region in the early Universe.

Looking to the future, this interpretation of $k_{\rm min}$ may offer clues
concerning how to extend our current semi-classical description of the early
Universe to scales below the Planck length, thereby heralding the initiation
of an observationally-motivated quantum gravity theory. In concert 
with such ideas, we point out that, if the $R_{\rm h}=ct$ cosmology turns out
to be correct, the potential of the (non-inflationary) scalar field present 
just after the Big Bang is precisely known (Eq.~1), allowing us eventually to 
also focus more directly on possible extensions to the standard model of particle
physics.

\section*{Acknowledgments}
I am grateful to the anonymous referee for his/her comments and 
suggestions, which have improved the presentation in this paper. I also 
acknowledge Amherst College for its support through a John Woodruff Simpson 
Fellowship.

\subsection*{Author contributions}

Fulvio Melia is the sole author of this paper.

\subsection*{Financial disclosure}

None reported.

\subsection*{Conflict of interest}

The authors declare no potential conflict of interests.

\bibliography{ms}%

\begin{thebibliography}{}

\bibitem [\protect \citeauthoryear {%
{Aad}%
\ \protect \BOthers {.}}{%
{Aad}%
\ \protect \BOthers {.}}{%
{\protect \APACyear {2012}}%
}]{%
Aad:2012}
\APACinsertmetastar {%
Aad:2012}%
\begin{APACrefauthors}%
{Aad}, G.%
, {Abajyan}, T.%
, {Abbott}, B.%
\ et al.\end{APACrefauthors}%
\unskip\
\newblock
\APACrefYearMonthDay{2012}{{\APACmonth{09}}}{},
\newblock
\unskip
\newblock
\APACjournalVolNumPages{Physics Letters B}{716}{1}{1-29}.
\newblock
\begin{APACrefDOI} \doi{10.1016/j.physletb.2012.08.020} \end{APACrefDOI}
\PrintBackRefs{\CurrentBib}

\bibitem [\protect \citeauthoryear {%
{Abbott}%
\ \BBA {} {Wise}%
}{%
{Abbott}%
\ \BBA {} {Wise}%
}{%
{\protect \APACyear {1984}}%
}]{%
Abbott:1984}
\APACinsertmetastar {%
Abbott:1984}%
\begin{APACrefauthors}%
{Abbott}, L\BPBI F.%
\BCBT {}\ \BBA {} {Wise}, M\BPBI B.%
\end{APACrefauthors}%
\unskip\
\newblock
\APACrefYearMonthDay{1984}{{\APACmonth{10}}}{},
\newblock
\unskip
\newblock
\APACjournalVolNumPages{Nuclear Physics B}{244}{2}{541-548}.
\newblock
\begin{APACrefDOI} \doi{10.1016/0550-3213(84)90329-8} \end{APACrefDOI}
\PrintBackRefs{\CurrentBib}

\bibitem [\protect \citeauthoryear {%
{Ashoorioon}%
, {Chialva}%
\BCBL {}\ \BBA {} {Danielsson}%
}{%
{Ashoorioon}%
\ \protect \BOthers {.}}{%
{\protect \APACyear {2011}}%
}]{%
Ashoorioon:2011}
\APACinsertmetastar {%
Ashoorioon:2011}%
\begin{APACrefauthors}%
{Ashoorioon}, A.%
, {Chialva}, D.%
\BCBL {}\ \BBA {} {Danielsson}, U.%
\end{APACrefauthors}%
\unskip\
\newblock
\APACrefYearMonthDay{2011}{{\APACmonth{06}}}{},
\newblock
\unskip
\newblock
\APACjournalVolNumPages{\jcap}{2011}{6}{034}.
\newblock
\begin{APACrefDOI} \doi{10.1088/1475-7516/2011/06/034} \end{APACrefDOI}
\PrintBackRefs{\CurrentBib}

\bibitem [\protect \citeauthoryear {%
{Ashtekar}%
, {Corichi}%
\BCBL {}\ \BBA {} {Singh}%
}{%
{Ashtekar}%
\ \protect \BOthers {.}}{%
{\protect \APACyear {2008}}%
}]{%
Ashtekar:2008}
\APACinsertmetastar {%
Ashtekar:2008}%
\begin{APACrefauthors}%
{Ashtekar}, A.%
, {Corichi}, A.%
\BCBL {}\ \BBA {} {Singh}, P.%
\end{APACrefauthors}%
\unskip\
\newblock
\APACrefYearMonthDay{2008}{{\APACmonth{01}}}{},
\newblock
\unskip
\newblock
\APACjournalVolNumPages{\prd}{77}{2}{024046}.
\newblock
\begin{APACrefDOI} \doi{10.1103/PhysRevD.77.024046} \end{APACrefDOI}
\PrintBackRefs{\CurrentBib}

\bibitem [\protect \citeauthoryear {%
{Barr}%
, {Dolan}%
, {Englert}%
, {de Lima}%
\BCBL {}\ \BBA {} {Spannowsky}%
}{%
{Barr}%
\ \protect \BOthers {.}}{%
{\protect \APACyear {2015}}%
}]{%
Barr:2015}
\APACinsertmetastar {%
Barr:2015}%
\begin{APACrefauthors}%
{Barr}, A\BPBI J.%
, {Dolan}, M\BPBI J.%
, {Englert}, C.%
, {de Lima}, D\BPBI E\BPBI F.%
\BCBL {}\ \BBA {} {Spannowsky}, M.%
\end{APACrefauthors}%
\unskip\
\newblock
\APACrefYearMonthDay{2015}{{\APACmonth{02}}}{},
\newblock
\unskip
\newblock
\APACjournalVolNumPages{Journal of High Energy Physics}{2015}{}{16}.
\newblock
\begin{APACrefDOI} \doi{10.1007/JHEP02(2015)016} \end{APACrefDOI}
\PrintBackRefs{\CurrentBib}

\bibitem [\protect \citeauthoryear {%
{Barrow}%
}{%
{Barrow}%
}{%
{\protect \APACyear {1987}}%
}]{%
Barrow:1987}
\APACinsertmetastar {%
Barrow:1987}%
\begin{APACrefauthors}%
{Barrow}, J\BPBI D.%
\end{APACrefauthors}%
\unskip\
\newblock
\APACrefYearMonthDay{1987}{{\APACmonth{03}}}{},
\newblock
\unskip
\newblock
\APACjournalVolNumPages{Physics Letters B}{187}{1-2}{12-16}.
\newblock
\begin{APACrefDOI} \doi{10.1016/0370-2693(87)90063-3} \end{APACrefDOI}
\PrintBackRefs{\CurrentBib}

\bibitem [\protect \citeauthoryear {%
{Bennett}%
\ \protect \BOthers {.}}{%
{Bennett}%
\ \protect \BOthers {.}}{%
{\protect \APACyear {2003}}%
}]{%
Bennett:2003}
\APACinsertmetastar {%
Bennett:2003}%
\begin{APACrefauthors}%
{Bennett}, C\BPBI L.%
, {Hill}, R\BPBI S.%
, {Hinshaw}, G.%
\ et al.\end{APACrefauthors}%
\unskip\
\newblock
\APACrefYearMonthDay{2003}{{\APACmonth{09}}}{},
\newblock
\unskip
\newblock
\APACjournalVolNumPages{\apjs}{148}{1}{97-117}.
\newblock
\begin{APACrefDOI} \doi{10.1086/377252} \end{APACrefDOI}
\PrintBackRefs{\CurrentBib}

\bibitem [\protect \citeauthoryear {%
R.~{Brandenberger}%
\ \BBA {} {Ho}%
}{%
R.~{Brandenberger}%
\ \BBA {} {Ho}%
}{%
{\protect \APACyear {2002}}%
}]{%
Brandenberger:2002}
\APACinsertmetastar {%
Brandenberger:2002}%
\begin{APACrefauthors}%
{Brandenberger}, R.%
\BCBT {}\ \BBA {} {Ho}, P\BHBI M.%
\end{APACrefauthors}%
\unskip\
\newblock
\APACrefYearMonthDay{2002}{{\APACmonth{07}}}{},
\newblock
\unskip
\newblock
\APACjournalVolNumPages{\prd}{66}{2}{023517}.
\newblock
\begin{APACrefDOI} \doi{10.1103/PhysRevD.66.023517} \end{APACrefDOI}
\PrintBackRefs{\CurrentBib}

\bibitem [\protect \citeauthoryear {%
R\BPBI H.~{Brandenberger}%
\ \BBA {} {Martin}%
}{%
R\BPBI H.~{Brandenberger}%
\ \BBA {} {Martin}%
}{%
{\protect \APACyear {2001}}%
}]{%
Brandenberger:2001}
\APACinsertmetastar {%
Brandenberger:2001}%
\begin{APACrefauthors}%
{Brandenberger}, R\BPBI H.%
\BCBT {}\ \BBA {} {Martin}, J.%
\end{APACrefauthors}%
\unskip\
\newblock
\APACrefYearMonthDay{2001}{{\APACmonth{01}}}{},
\newblock
\unskip
\newblock
\APACjournalVolNumPages{Modern Physics Letters A}{16}{15}{999-1006}.
\newblock
\begin{APACrefDOI} \doi{10.1142/S0217732301004170} \end{APACrefDOI}
\PrintBackRefs{\CurrentBib}

\bibitem [\protect \citeauthoryear {%
R\BPBI H.~{Brandenberger}%
\ \BBA {} {Martin}%
}{%
R\BPBI H.~{Brandenberger}%
\ \BBA {} {Martin}%
}{%
{\protect \APACyear {2002}}%
}]{%
Brandenberger:2002b}
\APACinsertmetastar {%
Brandenberger:2002b}%
\begin{APACrefauthors}%
{Brandenberger}, R\BPBI H.%
\BCBT {}\ \BBA {} {Martin}, J.%
\end{APACrefauthors}%
\unskip\
\newblock
\APACrefYearMonthDay{2002}{{\APACmonth{01}}}{},
\newblock
\unskip
\newblock
\APACjournalVolNumPages{International Journal of Modern Physics
  A}{17}{25}{3663-3680}.
\newblock
\begin{APACrefDOI} \doi{10.1142/S0217751X02010765} \end{APACrefDOI}
\PrintBackRefs{\CurrentBib}

\bibitem [\protect \citeauthoryear {%
R\BPBI H.~{Brandenberger}%
\ \BBA {} {Martin}%
}{%
R\BPBI H.~{Brandenberger}%
\ \BBA {} {Martin}%
}{%
{\protect \APACyear {2013}}%
}]{%
Brandenberger:2013}
\APACinsertmetastar {%
Brandenberger:2013}%
\begin{APACrefauthors}%
{Brandenberger}, R\BPBI H.%
\BCBT {}\ \BBA {} {Martin}, J.%
\end{APACrefauthors}%
\unskip\
\newblock
\APACrefYearMonthDay{2013}{{\APACmonth{06}}}{},
\newblock
\unskip
\newblock
\APACjournalVolNumPages{Classical and Quantum Gravity}{30}{11}{113001}.
\newblock
\begin{APACrefDOI} \doi{10.1088/0264-9381/30/11/113001} \end{APACrefDOI}
\PrintBackRefs{\CurrentBib}

\bibitem [\protect \citeauthoryear {%
{Bunch}%
\ \BBA {} {Davies}%
}{%
{Bunch}%
\ \BBA {} {Davies}%
}{%
{\protect \APACyear {1978}}%
}]{%
Bunch:1978}
\APACinsertmetastar {%
Bunch:1978}%
\begin{APACrefauthors}%
{Bunch}, T\BPBI S.%
\BCBT {}\ \BBA {} {Davies}, P\BPBI C\BPBI W.%
\end{APACrefauthors}%
\unskip\
\newblock
\APACrefYearMonthDay{1978}{{\APACmonth{03}}}{},
\newblock
\unskip
\newblock
\APACjournalVolNumPages{Proceedings of the Royal Society of London Series
  A}{360}{1700}{117-134}.
\newblock
\begin{APACrefDOI} \doi{10.1098/rspa.1978.0060} \end{APACrefDOI}
\PrintBackRefs{\CurrentBib}

\bibitem [\protect \citeauthoryear {%
{Chu}%
, {Greene}%
\BCBL {}\ \BBA {} {Shiu}%
}{%
{Chu}%
\ \protect \BOthers {.}}{%
{\protect \APACyear {2001}}%
}]{%
Chu:2001}
\APACinsertmetastar {%
Chu:2001}%
\begin{APACrefauthors}%
{Chu}, C\BHBI S.%
, {Greene}, B\BPBI R.%
\BCBL {}\ \BBA {} {Shiu}, G.%
\end{APACrefauthors}%
\unskip\
\newblock
\APACrefYearMonthDay{2001}{{\APACmonth{01}}}{},
\newblock
\unskip
\newblock
\APACjournalVolNumPages{Modern Physics Letters A}{16}{34}{2231-2240}.
\newblock
\begin{APACrefDOI} \doi{10.1142/S0217732301005680} \end{APACrefDOI}
\PrintBackRefs{\CurrentBib}

\bibitem [\protect \citeauthoryear {%
{Collins}%
\ \BBA {} {Holman}%
}{%
{Collins}%
\ \BBA {} {Holman}%
}{%
{\protect \APACyear {2005}}%
}]{%
Collins:2005}
\APACinsertmetastar {%
Collins:2005}%
\begin{APACrefauthors}%
{Collins}, H.%
\BCBT {}\ \BBA {} {Holman}, R.%
\end{APACrefauthors}%
\unskip\
\newblock
\APACrefYearMonthDay{2005}{{\APACmonth{04}}}{},
\newblock
\unskip
\newblock
\APACjournalVolNumPages{\prd}{71}{8}{085009}.
\newblock
\begin{APACrefDOI} \doi{10.1103/PhysRevD.71.085009} \end{APACrefDOI}
\PrintBackRefs{\CurrentBib}

\bibitem [\protect \citeauthoryear {%
{Danielsson}%
}{%
{Danielsson}%
}{%
{\protect \APACyear {2002}}%
}]{%
Danielsson:2002}
\APACinsertmetastar {%
Danielsson:2002}%
\begin{APACrefauthors}%
{Danielsson}, U\BPBI H.%
\end{APACrefauthors}%
\unskip\
\newblock
\APACrefYearMonthDay{2002}{{\APACmonth{07}}}{},
\newblock
\unskip
\newblock
\APACjournalVolNumPages{\prd}{66}{2}{023511}.
\newblock
\begin{APACrefDOI} \doi{10.1103/PhysRevD.66.023511} \end{APACrefDOI}
\PrintBackRefs{\CurrentBib}

\bibitem [\protect \citeauthoryear {%
{Dewitt}%
}{%
{Dewitt}%
}{%
{\protect \APACyear {1967}}%
}]{%
DeWitt:1967}
\APACinsertmetastar {%
DeWitt:1967}%
\begin{APACrefauthors}%
{Dewitt}, B\BPBI S.%
\end{APACrefauthors}%
\unskip\
\newblock
\APACrefYearMonthDay{1967}{{\APACmonth{08}}}{},
\newblock
\unskip
\newblock
\APACjournalVolNumPages{Physical Review}{160}{5}{1113-1148}.
\newblock
\begin{APACrefDOI} \doi{10.1103/PhysRev.160.1113} \end{APACrefDOI}
\PrintBackRefs{\CurrentBib}

\bibitem [\protect \citeauthoryear {%
{Dolan}%
, {Englert}%
\BCBL {}\ \BBA {} {Spannowsky}%
}{%
{Dolan}%
\ \protect \BOthers {.}}{%
{\protect \APACyear {2012}}%
}]{%
Dolan:2012}
\APACinsertmetastar {%
Dolan:2012}%
\begin{APACrefauthors}%
{Dolan}, M\BPBI J.%
, {Englert}, C.%
\BCBL {}\ \BBA {} {Spannowsky}, M.%
\end{APACrefauthors}%
\unskip\
\newblock
\APACrefYearMonthDay{2012}{{\APACmonth{10}}}{},
\newblock
\unskip
\newblock
\APACjournalVolNumPages{Journal of High Energy Physics}{2012}{}{112}.
\newblock
\begin{APACrefDOI} \doi{10.1007/JHEP10(2012)112} \end{APACrefDOI}
\PrintBackRefs{\CurrentBib}

\bibitem [\protect \citeauthoryear {%
{Easther}%
, {Greene}%
, {Kinney}%
\BCBL {}\ \BBA {} {Shiu}%
}{%
{Easther}%
\ \protect \BOthers {.}}{%
{\protect \APACyear {2001}}%
}]{%
Easther:2001}
\APACinsertmetastar {%
Easther:2001}%
\begin{APACrefauthors}%
{Easther}, R.%
, {Greene}, B\BPBI R.%
, {Kinney}, W\BPBI H.%
\BCBL {}\ \BBA {} {Shiu}, G.%
\end{APACrefauthors}%
\unskip\
\newblock
\APACrefYearMonthDay{2001}{{\APACmonth{11}}}{},
\newblock
\unskip
\newblock
\APACjournalVolNumPages{\prd}{64}{10}{103502}.
\newblock
\begin{APACrefDOI} \doi{10.1103/PhysRevD.64.103502} \end{APACrefDOI}
\PrintBackRefs{\CurrentBib}

\bibitem [\protect \citeauthoryear {%
{Easther}%
, {Greene}%
, {Kinney}%
\BCBL {}\ \BBA {} {Shiu}%
}{%
{Easther}%
\ \protect \BOthers {.}}{%
{\protect \APACyear {2002}}%
}]{%
Easther:2002}
\APACinsertmetastar {%
Easther:2002}%
\begin{APACrefauthors}%
{Easther}, R.%
, {Greene}, B\BPBI R.%
, {Kinney}, W\BPBI H.%
\BCBL {}\ \BBA {} {Shiu}, G.%
\end{APACrefauthors}%
\unskip\
\newblock
\APACrefYearMonthDay{2002}{{\APACmonth{07}}}{},
\newblock
\unskip
\newblock
\APACjournalVolNumPages{\prd}{66}{2}{023518}.
\newblock
\begin{APACrefDOI} \doi{10.1103/PhysRevD.66.023518} \end{APACrefDOI}
\PrintBackRefs{\CurrentBib}

\bibitem [\protect \citeauthoryear {%
{Fileviez P{\'e}rez}%
, {Patel}%
, {Ramsey-Musolf}%
\BCBL {}\ \BBA {} {Wang}%
}{%
{Fileviez P{\'e}rez}%
\ \protect \BOthers {.}}{%
{\protect \APACyear {2009}}%
}]{%
Fileviez:2009}
\APACinsertmetastar {%
Fileviez:2009}%
\begin{APACrefauthors}%
{Fileviez P{\'e}rez}, P.%
, {Patel}, H\BPBI H.%
, {Ramsey-Musolf}, M\BPBI J.%
\BCBL {}\ \BBA {} {Wang}, K.%
\end{APACrefauthors}%
\unskip\
\newblock
\APACrefYearMonthDay{2009}{{\APACmonth{03}}}{},
\newblock
\unskip
\newblock
\APACjournalVolNumPages{\prd}{79}{5}{055024}.
\newblock
\begin{APACrefDOI} \doi{10.1103/PhysRevD.79.055024} \end{APACrefDOI}
\PrintBackRefs{\CurrentBib}

\bibitem [\protect \citeauthoryear {%
{Guth}%
}{%
{Guth}%
}{%
{\protect \APACyear {1981}}%
}]{%
Guth:1981}
\APACinsertmetastar {%
Guth:1981}%
\begin{APACrefauthors}%
{Guth}, A\BPBI H.%
\end{APACrefauthors}%
\unskip\
\newblock
\APACrefYearMonthDay{1981}{{\APACmonth{01}}}{},
\newblock
\unskip
\newblock
\APACjournalVolNumPages{\prd}{23}{2}{347-356}.
\newblock
\begin{APACrefDOI} \doi{10.1103/PhysRevD.23.347} \end{APACrefDOI}
\PrintBackRefs{\CurrentBib}

\bibitem [\protect \citeauthoryear {%
{Hassan}%
\ \BBA {} {Sloth}%
}{%
{Hassan}%
\ \BBA {} {Sloth}%
}{%
{\protect \APACyear {2003}}%
}]{%
Hassan:2003}
\APACinsertmetastar {%
Hassan:2003}%
\begin{APACrefauthors}%
{Hassan}, S\BPBI F.%
\BCBT {}\ \BBA {} {Sloth}, M\BPBI S.%
\end{APACrefauthors}%
\unskip\
\newblock
\APACrefYearMonthDay{2003}{{\APACmonth{12}}}{},
\newblock
\unskip
\newblock
\APACjournalVolNumPages{Nuclear Physics B}{674}{1-2}{434-458}.
\newblock
\begin{APACrefDOI} \doi{10.1016/j.nuclphysb.2003.09.041} \end{APACrefDOI}
\PrintBackRefs{\CurrentBib}

\bibitem [\protect \citeauthoryear {%
{Hinshaw}%
\ \protect \BOthers {.}}{%
{Hinshaw}%
\ \protect \BOthers {.}}{%
{\protect \APACyear {1996}}%
}]{%
Hinshaw:1996}
\APACinsertmetastar {%
Hinshaw:1996}%
\begin{APACrefauthors}%
{Hinshaw}, G.%
, {Branday}, A\BPBI J.%
, {Bennett}, C\BPBI L.%
\ et al.\end{APACrefauthors}%
\unskip\
\newblock
\APACrefYearMonthDay{1996}{{\APACmonth{06}}}{},
\newblock
\unskip
\newblock
\APACjournalVolNumPages{\apjl}{464}{}{L25}.
\newblock
\begin{APACrefDOI} \doi{10.1086/310076} \end{APACrefDOI}
\PrintBackRefs{\CurrentBib}

\bibitem [\protect \citeauthoryear {%
{Hollands}%
\ \BBA {} {Wald}%
}{%
{Hollands}%
\ \BBA {} {Wald}%
}{%
{\protect \APACyear {2002}}%
}]{%
Hollands:2002}
\APACinsertmetastar {%
Hollands:2002}%
\begin{APACrefauthors}%
{Hollands}, S.%
\BCBT {}\ \BBA {} {Wald}, R\BPBI M.%
\end{APACrefauthors}%
\unskip\
\newblock
\APACrefYearMonthDay{2002}{{\APACmonth{12}}}{},
\newblock
\unskip
\newblock
\APACjournalVolNumPages{General Relativity and Gravitation}{34}{12}{2043-2055}.
\newblock
\begin{APACrefDOI} \doi{10.1023/A:1021175216055} \end{APACrefDOI}
\PrintBackRefs{\CurrentBib}

\bibitem [\protect \citeauthoryear {%
{Ijjas}%
, {Steinhardt}%
\BCBL {}\ \BBA {} {Loeb}%
}{%
{Ijjas}%
\ \protect \BOthers {.}}{%
{\protect \APACyear {2013}}%
}]{%
Ijjas:2013}
\APACinsertmetastar {%
Ijjas:2013}%
\begin{APACrefauthors}%
{Ijjas}, A.%
, {Steinhardt}, P\BPBI J.%
\BCBL {}\ \BBA {} {Loeb}, A.%
\end{APACrefauthors}%
\unskip\
\newblock
\APACrefYearMonthDay{2013}{{\APACmonth{06}}}{},
\newblock
\unskip
\newblock
\APACjournalVolNumPages{Physics Letters B}{723}{4-5}{261-266}.
\newblock
\begin{APACrefDOI} \doi{10.1016/j.physletb.2013.05.023} \end{APACrefDOI}
\PrintBackRefs{\CurrentBib}

\bibitem [\protect \citeauthoryear {%
{Jor{\'a}s}%
\ \BBA {} {Marozzi}%
}{%
{Jor{\'a}s}%
\ \BBA {} {Marozzi}%
}{%
{\protect \APACyear {2009}}%
}]{%
Joras:2009}
\APACinsertmetastar {%
Joras:2009}%
\begin{APACrefauthors}%
{Jor{\'a}s}, S\BPBI E.%
\BCBT {}\ \BBA {} {Marozzi}, G.%
\end{APACrefauthors}%
\unskip\
\newblock
\APACrefYearMonthDay{2009}{{\APACmonth{01}}}{},
\newblock
\unskip
\newblock
\APACjournalVolNumPages{\prd}{79}{2}{023514}.
\newblock
\begin{APACrefDOI} \doi{10.1103/PhysRevD.79.023514} \end{APACrefDOI}
\PrintBackRefs{\CurrentBib}

\bibitem [\protect \citeauthoryear {%
{Kaloper}%
, {Kleban}%
, {Lawrence}%
\BCBL {}\ \BBA {} {Shenker}%
}{%
{Kaloper}%
\ \protect \BOthers {.}}{%
{\protect \APACyear {2002}}%
}]{%
Kaloper:2002}
\APACinsertmetastar {%
Kaloper:2002}%
\begin{APACrefauthors}%
{Kaloper}, N.%
, {Kleban}, M.%
, {Lawrence}, A.%
\BCBL {}\ \BBA {} {Shenker}, S.%
\end{APACrefauthors}%
\unskip\
\newblock
\APACrefYearMonthDay{2002}{{\APACmonth{12}}}{},
\newblock
\unskip
\newblock
\APACjournalVolNumPages{\prd}{66}{12}{123510}.
\newblock
\begin{APACrefDOI} \doi{10.1103/PhysRevD.66.123510} \end{APACrefDOI}
\PrintBackRefs{\CurrentBib}

\bibitem [\protect \citeauthoryear {%
{Kempf}%
}{%
{Kempf}%
}{%
{\protect \APACyear {2001}}%
}]{%
Kempf:2001}
\APACinsertmetastar {%
Kempf:2001}%
\begin{APACrefauthors}%
{Kempf}, A.%
\end{APACrefauthors}%
\unskip\
\newblock
\APACrefYearMonthDay{2001}{{\APACmonth{04}}}{},
\newblock
\unskip
\newblock
\APACjournalVolNumPages{\prd}{63}{8}{083514}.
\newblock
\begin{APACrefDOI} \doi{10.1103/PhysRevD.63.083514} \end{APACrefDOI}
\PrintBackRefs{\CurrentBib}

\bibitem [\protect \citeauthoryear {%
{Kempf}%
\ \BBA {} {Niemeyer}%
}{%
{Kempf}%
\ \BBA {} {Niemeyer}%
}{%
{\protect \APACyear {2001}}%
}]{%
Kempf:2001b}
\APACinsertmetastar {%
Kempf:2001b}%
\begin{APACrefauthors}%
{Kempf}, A.%
\BCBT {}\ \BBA {} {Niemeyer}, J\BPBI C.%
\end{APACrefauthors}%
\unskip\
\newblock
\APACrefYearMonthDay{2001}{{\APACmonth{11}}}{},
\newblock
\unskip
\newblock
\APACjournalVolNumPages{\prd}{64}{10}{103501}.
\newblock
\begin{APACrefDOI} \doi{10.1103/PhysRevD.64.103501} \end{APACrefDOI}
\PrintBackRefs{\CurrentBib}

\bibitem [\protect \citeauthoryear {%
{Kiefer}%
}{%
{Kiefer}%
}{%
{\protect \APACyear {2004}}%
}]{%
Kiefer:2004}
\APACinsertmetastar {%
Kiefer:2004}%
\begin{APACrefauthors}%
{Kiefer}, C.%
\end{APACrefauthors}%
\unskip\
\newblock
\APACrefYearMonthDay{2004}{}{},
\newblock
{\BBOQ}\APACrefatitle {{Quantum Gravity and Fundamental Constants}} {{Quantum
  Gravity and Fundamental Constants}}.{\BBCQ}
\newblock
\BIn{} S\BPBI G.~{Karshenboim}\ \BBA {} E.~{Peik}\ (\BEDS), \APACrefbtitle
  {Astrophysics, Clocks and Fundamental Constants} {Astrophysics, Clocks and
  Fundamental Constants}\ \BVOL~648, \BPG~115-127.
\newblock
\begin{APACrefDOI} \doi{10.1007/978-3-540-40991-5_8} \end{APACrefDOI}
\PrintBackRefs{\CurrentBib}

\bibitem [\protect \citeauthoryear {%
{Kodama}%
\ \BBA {} {Sasaki}%
}{%
{Kodama}%
\ \BBA {} {Sasaki}%
}{%
{\protect \APACyear {1984}}%
}]{%
Kodama:1984}
\APACinsertmetastar {%
Kodama:1984}%
\begin{APACrefauthors}%
{Kodama}, H.%
\BCBT {}\ \BBA {} {Sasaki}, M.%
\end{APACrefauthors}%
\unskip\
\newblock
\APACrefYearMonthDay{1984}{{\APACmonth{01}}}{},
\newblock
\unskip
\newblock
\APACjournalVolNumPages{Progress of Theoretical Physics Supplement}{78}{}{1}.
\newblock
\begin{APACrefDOI} \doi{10.1143/PTPS.78.1} \end{APACrefDOI}
\PrintBackRefs{\CurrentBib}

\bibitem [\protect \citeauthoryear {%
{Kucha{\v{r}}}%
}{%
{Kucha{\v{r}}}%
}{%
{\protect \APACyear {2011}}%
}]{%
Kuchar:2011}
\APACinsertmetastar {%
Kuchar:2011}%
\begin{APACrefauthors}%
{Kucha{\v{r}}}, K\BPBI V.%
\end{APACrefauthors}%
\unskip\
\newblock
\APACrefYearMonthDay{2011}{{\APACmonth{01}}}{},
\newblock
\unskip
\newblock
\APACjournalVolNumPages{International Journal of Modern Physics D}{20}{}{3-86}.
\newblock
\begin{APACrefDOI} \doi{10.1142/S0218271811019347} \end{APACrefDOI}
\PrintBackRefs{\CurrentBib}

\bibitem [\protect \citeauthoryear {%
{Liddle}%
}{%
{Liddle}%
}{%
{\protect \APACyear {1989}}%
}]{%
Liddle:1989}
\APACinsertmetastar {%
Liddle:1989}%
\begin{APACrefauthors}%
{Liddle}, A\BPBI R.%
\end{APACrefauthors}%
\unskip\
\newblock
\APACrefYearMonthDay{1989}{{\APACmonth{04}}}{},
\newblock
\unskip
\newblock
\APACjournalVolNumPages{Physics Letters B}{220}{4}{502-508}.
\newblock
\begin{APACrefDOI} \doi{10.1016/0370-2693(89)90776-4} \end{APACrefDOI}
\PrintBackRefs{\CurrentBib}

\bibitem [\protect \citeauthoryear {%
{Linde}%
}{%
{Linde}%
}{%
{\protect \APACyear {1982}}%
}]{%
Linde:1982}
\APACinsertmetastar {%
Linde:1982}%
\begin{APACrefauthors}%
{Linde}, A\BPBI D.%
\end{APACrefauthors}%
\unskip\
\newblock
\APACrefYearMonthDay{1982}{{\APACmonth{02}}}{},
\newblock
\unskip
\newblock
\APACjournalVolNumPages{Physics Letters B}{108}{6}{389-393}.
\newblock
\begin{APACrefDOI} \doi{10.1016/0370-2693(82)91219-9} \end{APACrefDOI}
\PrintBackRefs{\CurrentBib}

\bibitem [\protect \citeauthoryear {%
{Liu}%
\ \BBA {} {Melia}%
}{%
{Liu}%
\ \BBA {} {Melia}%
}{%
{\protect \APACyear {2020}}%
}]{%
LiuMelia:2020}
\APACinsertmetastar {%
LiuMelia:2020}%
\begin{APACrefauthors}%
{Liu}, J.%
\BCBT {}\ \BBA {} {Melia}, F.%
\end{APACrefauthors}%
\unskip\
\newblock
\APACrefYearMonthDay{2020}{{\APACmonth{06}}}{},
\newblock
\unskip
\newblock
\APACjournalVolNumPages{arXiv e-prints}{}{}{arXiv:2006.02510}.
\PrintBackRefs{\CurrentBib}

\bibitem [\protect \citeauthoryear {%
{Lizzi}%
, {Mangano}%
, {Miele}%
\BCBL {}\ \BBA {} {Peloso}%
}{%
{Lizzi}%
\ \protect \BOthers {.}}{%
{\protect \APACyear {2002}}%
}]{%
Lizzi:2002}
\APACinsertmetastar {%
Lizzi:2002}%
\begin{APACrefauthors}%
{Lizzi}, F.%
, {Mangano}, G.%
, {Miele}, G.%
\BCBL {}\ \BBA {} {Peloso}, M.%
\end{APACrefauthors}%
\unskip\
\newblock
\APACrefYearMonthDay{2002}{{\APACmonth{06}}}{},
\newblock
\unskip
\newblock
\APACjournalVolNumPages{Journal of High Energy Physics}{2002}{6}{049}.
\newblock
\begin{APACrefDOI} \doi{10.1088/1126-6708/2002/06/049} \end{APACrefDOI}
\PrintBackRefs{\CurrentBib}

\bibitem [\protect \citeauthoryear {%
{Lucchin}%
\ \BBA {} {Matarrese}%
}{%
{Lucchin}%
\ \BBA {} {Matarrese}%
}{%
{\protect \APACyear {1985}}%
}]{%
Lucchin:1985}
\APACinsertmetastar {%
Lucchin:1985}%
\begin{APACrefauthors}%
{Lucchin}, F.%
\BCBT {}\ \BBA {} {Matarrese}, S.%
\end{APACrefauthors}%
\unskip\
\newblock
\APACrefYearMonthDay{1985}{{\APACmonth{09}}}{},
\newblock
\unskip
\newblock
\APACjournalVolNumPages{\prd}{32}{6}{1316-1322}.
\newblock
\begin{APACrefDOI} \doi{10.1103/PhysRevD.32.1316} \end{APACrefDOI}
\PrintBackRefs{\CurrentBib}

\bibitem [\protect \citeauthoryear {%
{Maldacena}%
}{%
{Maldacena}%
}{%
{\protect \APACyear {2003}}%
}]{%
Maldacena:2003}
\APACinsertmetastar {%
Maldacena:2003}%
\begin{APACrefauthors}%
{Maldacena}, J.%
\end{APACrefauthors}%
\unskip\
\newblock
\APACrefYearMonthDay{2003}{{\APACmonth{05}}}{},
\newblock
\unskip
\newblock
\APACjournalVolNumPages{Journal of High Energy Physics}{2003}{5}{013}.
\newblock
\begin{APACrefDOI} \doi{10.1088/1126-6708/2003/05/013} \end{APACrefDOI}
\PrintBackRefs{\CurrentBib}

\bibitem [\protect \citeauthoryear {%
{Martin}%
\ \BBA {} {Brandenberger}%
}{%
{Martin}%
\ \BBA {} {Brandenberger}%
}{%
{\protect \APACyear {2001}}%
}]{%
Martin:2001}
\APACinsertmetastar {%
Martin:2001}%
\begin{APACrefauthors}%
{Martin}, J.%
\BCBT {}\ \BBA {} {Brandenberger}, R\BPBI H.%
\end{APACrefauthors}%
\unskip\
\newblock
\APACrefYearMonthDay{2001}{{\APACmonth{06}}}{},
\newblock
\unskip
\newblock
\APACjournalVolNumPages{\prd}{63}{12}{123501}.
\newblock
\begin{APACrefDOI} \doi{10.1103/PhysRevD.63.123501} \end{APACrefDOI}
\PrintBackRefs{\CurrentBib}

\bibitem [\protect \citeauthoryear {%
{Melia}%
}{%
{Melia}%
}{%
{\protect \APACyear {2007}}%
}]{%
Melia:2007}
\APACinsertmetastar {%
Melia:2007}%
\begin{APACrefauthors}%
{Melia}, F.%
\end{APACrefauthors}%
\unskip\
\newblock
\APACrefYearMonthDay{2007}{{\APACmonth{12}}}{},
\newblock
\unskip
\newblock
\APACjournalVolNumPages{\mnras}{382}{4}{1917-1921}.
\newblock
\begin{APACrefDOI} \doi{10.1111/j.1365-2966.2007.12499.x} \end{APACrefDOI}
\PrintBackRefs{\CurrentBib}

\bibitem [\protect \citeauthoryear {%
{Melia}%
}{%
{Melia}%
}{%
{\protect \APACyear {2013}}%
}]{%
Melia:2013a}
\APACinsertmetastar {%
Melia:2013a}%
\begin{APACrefauthors}%
{Melia}, F.%
\end{APACrefauthors}%
\unskip\
\newblock
\APACrefYearMonthDay{2013}{{\APACmonth{05}}}{},
\newblock
\unskip
\newblock
\APACjournalVolNumPages{\aap}{553}{}{A76}.
\newblock
\begin{APACrefDOI} \doi{10.1051/0004-6361/201220447} \end{APACrefDOI}
\PrintBackRefs{\CurrentBib}

\bibitem [\protect \citeauthoryear {%
{Melia}%
}{%
{Melia}%
}{%
{\protect \APACyear {2016}}%
}]{%
Melia:2016a}
\APACinsertmetastar {%
Melia:2016a}%
\begin{APACrefauthors}%
{Melia}, F.%
\end{APACrefauthors}%
\unskip\
\newblock
\APACrefYearMonthDay{2016}{{\APACmonth{08}}}{},
\newblock
\unskip
\newblock
\APACjournalVolNumPages{Frontiers of Physics}{11}{4}{119801}.
\newblock
\begin{APACrefDOI} \doi{10.1007/s11467-016-0557-6} \end{APACrefDOI}
\PrintBackRefs{\CurrentBib}

\bibitem [\protect \citeauthoryear {%
{Melia}%
}{%
{Melia}%
}{%
{\protect \APACyear {2017}}%
{\protect \APACexlab {{\protect \BCnt {1}}}}}]{%
Melia:2017b}
\APACinsertmetastar {%
Melia:2017b}%
\begin{APACrefauthors}%
{Melia}, F.%
\end{APACrefauthors}%
\unskip\
\newblock
\APACrefYearMonthDay{2017{\protect \BCnt {1}}}{{\APACmonth{01}}}{},
\newblock
\unskip
\newblock
\APACjournalVolNumPages{Classical and Quantum Gravity}{34}{1}{015011}.
\newblock
\begin{APACrefDOI} \doi{10.1088/1361-6382/34/1/015011} \end{APACrefDOI}
\PrintBackRefs{\CurrentBib}

\bibitem [\protect \citeauthoryear {%
{Melia}%
}{%
{Melia}%
}{%
{\protect \APACyear {2017}}%
{\protect \APACexlab {{\protect \BCnt {2}}}}}]{%
Melia:2017a}
\APACinsertmetastar {%
Melia:2017a}%
\begin{APACrefauthors}%
{Melia}, F.%
\end{APACrefauthors}%
\unskip\
\newblock
\APACrefYearMonthDay{2017{\protect \BCnt {2}}}{{\APACmonth{02}}}{},
\newblock
\unskip
\newblock
\APACjournalVolNumPages{Frontiers of Physics}{12}{1}{129802}.
\newblock
\begin{APACrefDOI} \doi{10.1007/s11467-016-0611-4} \end{APACrefDOI}
\PrintBackRefs{\CurrentBib}

\bibitem [\protect \citeauthoryear {%
{Melia}%
}{%
{Melia}%
}{%
{\protect \APACyear {2018}}%
{\protect \APACexlab {{\protect \BCnt {1}}}}}]{%
Melia:2018c}
\APACinsertmetastar {%
Melia:2018c}%
\begin{APACrefauthors}%
{Melia}, F.%
\end{APACrefauthors}%
\unskip\
\newblock
\APACrefYearMonthDay{2018{\protect \BCnt {1}}}{{\APACmonth{08}}}{},
\newblock
\unskip
\newblock
\APACjournalVolNumPages{American Journal of Physics}{86}{8}{585-593}.
\newblock
\begin{APACrefDOI} \doi{10.1119/1.5045333} \end{APACrefDOI}
\PrintBackRefs{\CurrentBib}

\bibitem [\protect \citeauthoryear {%
{Melia}%
}{%
{Melia}%
}{%
{\protect \APACyear {2018}}%
{\protect \APACexlab {{\protect \BCnt {2}}}}}]{%
Melia:2018b}
\APACinsertmetastar {%
Melia:2018b}%
\begin{APACrefauthors}%
{Melia}, F.%
\end{APACrefauthors}%
\unskip\
\newblock
\APACrefYearMonthDay{2018{\protect \BCnt {2}}}{{\APACmonth{12}}}{},
\newblock
\unskip
\newblock
\APACjournalVolNumPages{\mnras}{481}{4}{4855-4862}.
\newblock
\begin{APACrefDOI} \doi{10.1093/mnras/sty2596} \end{APACrefDOI}
\PrintBackRefs{\CurrentBib}

\bibitem [\protect \citeauthoryear {%
{Melia}%
}{%
{Melia}%
}{%
{\protect \APACyear {2018}}%
{\protect \APACexlab {{\protect \BCnt {3}}}}}]{%
Melia:2018a}
\APACinsertmetastar {%
Melia:2018a}%
\begin{APACrefauthors}%
{Melia}, F.%
\end{APACrefauthors}%
\unskip\
\newblock
\APACrefYearMonthDay{2018{\protect \BCnt {3}}}{{\APACmonth{09}}}{},
\newblock
\unskip
\newblock
\APACjournalVolNumPages{European Physical Journal C}{78}{9}{739}.
\newblock
\begin{APACrefDOI} \doi{10.1140/epjc/s10052-018-6231-0} \end{APACrefDOI}
\PrintBackRefs{\CurrentBib}

\bibitem [\protect \citeauthoryear {%
{Melia}%
}{%
{Melia}%
}{%
{\protect \APACyear {2019}}%
{\protect \APACexlab {{\protect \BCnt {1}}}}}]{%
Melia:2019b}
\APACinsertmetastar {%
Melia:2019b}%
\begin{APACrefauthors}%
{Melia}, F.%
\end{APACrefauthors}%
\unskip\
\newblock
\APACrefYearMonthDay{2019{\protect \BCnt {1}}}{{\APACmonth{12}}}{},
\newblock
\unskip
\newblock
\APACjournalVolNumPages{Annals of Physics}{411}{}{167997}.
\newblock
\begin{APACrefDOI} \doi{10.1016/j.aop.2019.167997} \end{APACrefDOI}
\PrintBackRefs{\CurrentBib}

\bibitem [\protect \citeauthoryear {%
{Melia}%
}{%
{Melia}%
}{%
{\protect \APACyear {2019}}%
{\protect \APACexlab {{\protect \BCnt {2}}}}}]{%
Melia:2019a}
\APACinsertmetastar {%
Melia:2019a}%
\begin{APACrefauthors}%
{Melia}, F.%
\end{APACrefauthors}%
\unskip\
\newblock
\APACrefYearMonthDay{2019{\protect \BCnt {2}}}{{\APACmonth{08}}}{},
\newblock
\unskip
\newblock
\APACjournalVolNumPages{Modern Physics Letters A}{34}{26}{1930004-30}.
\newblock
\begin{APACrefDOI} \doi{10.1142/S0217732319300040} \end{APACrefDOI}
\PrintBackRefs{\CurrentBib}

\bibitem [\protect \citeauthoryear {%
Melia%
}{%
Melia%
}{%
{\protect \APACyear {2020}}%
}]{%
Melia:2020}
\APACinsertmetastar {%
Melia:2020}%
\begin{APACrefauthors}%
Melia, F.%
\end{APACrefauthors}%
\unskip\
\newblock
\APACrefYear{2020},
\newblock
\APACrefbtitle {The Cosmic Spacetime} {The Cosmic Spacetime}\
  (\PrintOrdinal{1}\ \BEd).
\newblock
\APACaddressPublisher{Oxfordshire, UK}{Taylor \& Francis}.
\PrintBackRefs{\CurrentBib}

\bibitem [\protect \citeauthoryear {%
{Melia}%
\ \BBA {} {Abdelqader}%
}{%
{Melia}%
\ \BBA {} {Abdelqader}%
}{%
{\protect \APACyear {2009}}%
}]{%
MeliaAbdelqader:2009}
\APACinsertmetastar {%
MeliaAbdelqader:2009}%
\begin{APACrefauthors}%
{Melia}, F.%
\BCBT {}\ \BBA {} {Abdelqader}, M.%
\end{APACrefauthors}%
\unskip\
\newblock
\APACrefYearMonthDay{2009}{{\APACmonth{01}}}{},
\newblock
\unskip
\newblock
\APACjournalVolNumPages{International Journal of Modern Physics
  D}{18}{12}{1889-1901}.
\newblock
\begin{APACrefDOI} \doi{10.1142/S0218271809015746} \end{APACrefDOI}
\PrintBackRefs{\CurrentBib}

\bibitem [\protect \citeauthoryear {%
{Melia}%
\ \BBA {} {L{\'o}pez-Corredoira}%
}{%
{Melia}%
\ \BBA {} {L{\'o}pez-Corredoira}%
}{%
{\protect \APACyear {2018}}%
}]{%
MeliaLopez:2018}
\APACinsertmetastar {%
MeliaLopez:2018}%
\begin{APACrefauthors}%
{Melia}, F.%
\BCBT {}\ \BBA {} {L{\'o}pez-Corredoira}, M.%
\end{APACrefauthors}%
\unskip\
\newblock
\APACrefYearMonthDay{2018}{{\APACmonth{03}}}{},
\newblock
\unskip
\newblock
\APACjournalVolNumPages{\aap}{610}{}{A87}.
\newblock
\begin{APACrefDOI} \doi{10.1051/0004-6361/201732181} \end{APACrefDOI}
\PrintBackRefs{\CurrentBib}

\bibitem [\protect \citeauthoryear {%
{Melia}%
\ \BBA {} {Shevchuk}%
}{%
{Melia}%
\ \BBA {} {Shevchuk}%
}{%
{\protect \APACyear {2012}}%
}]{%
MeliaShevchuk:2012}
\APACinsertmetastar {%
MeliaShevchuk:2012}%
\begin{APACrefauthors}%
{Melia}, F.%
\BCBT {}\ \BBA {} {Shevchuk}, A\BPBI S\BPBI H.%
\end{APACrefauthors}%
\unskip\
\newblock
\APACrefYearMonthDay{2012}{{\APACmonth{01}}}{},
\newblock
\unskip
\newblock
\APACjournalVolNumPages{\mnras}{419}{3}{2579-2586}.
\newblock
\begin{APACrefDOI} \doi{10.1111/j.1365-2966.2011.19906.x} \end{APACrefDOI}
\PrintBackRefs{\CurrentBib}

\bibitem [\protect \citeauthoryear {%
V.~{Mukhanov}%
}{%
V.~{Mukhanov}%
}{%
{\protect \APACyear {2005}}%
}]{%
Mukhanov:2005}
\APACinsertmetastar {%
Mukhanov:2005}%
\begin{APACrefauthors}%
{Mukhanov}, V.%
\end{APACrefauthors}%
\unskip\
\newblock
\APACrefYear{2005},
\newblock
\APACrefbtitle {{Physical Foundations of Cosmology}} {{Physical Foundations of
  Cosmology}}.
\newblock
\begin{APACrefDOI} \doi{10.2277/0521563984} \end{APACrefDOI}
\PrintBackRefs{\CurrentBib}

\bibitem [\protect \citeauthoryear {%
V\BPBI F.~{Mukhanov}%
, {Feldman}%
\BCBL {}\ \BBA {} {Brandenberger}%
}{%
V\BPBI F.~{Mukhanov}%
\ \protect \BOthers {.}}{%
{\protect \APACyear {1992}}%
}]{%
Mukhanov:1992}
\APACinsertmetastar {%
Mukhanov:1992}%
\begin{APACrefauthors}%
{Mukhanov}, V\BPBI F.%
, {Feldman}, H\BPBI A.%
\BCBL {}\ \BBA {} {Brandenberger}, R\BPBI H.%
\end{APACrefauthors}%
\unskip\
\newblock
\APACrefYearMonthDay{1992}{{\APACmonth{06}}}{},
\newblock
\unskip
\newblock
\APACjournalVolNumPages{\physrep}{215}{5-6}{203-333}.
\newblock
\begin{APACrefDOI} \doi{10.1016/0370-1573(92)90044-Z} \end{APACrefDOI}
\PrintBackRefs{\CurrentBib}

\bibitem [\protect \citeauthoryear {%
{Niemeyer}%
}{%
{Niemeyer}%
}{%
{\protect \APACyear {2001}}%
}]{%
Niemeyer:2001}
\APACinsertmetastar {%
Niemeyer:2001}%
\begin{APACrefauthors}%
{Niemeyer}, J\BPBI C.%
\end{APACrefauthors}%
\unskip\
\newblock
\APACrefYearMonthDay{2001}{{\APACmonth{06}}}{},
\newblock
\unskip
\newblock
\APACjournalVolNumPages{\prd}{63}{12}{123502}.
\newblock
\begin{APACrefDOI} \doi{10.1103/PhysRevD.63.123502} \end{APACrefDOI}
\PrintBackRefs{\CurrentBib}

\bibitem [\protect \citeauthoryear {%
{Niemeyer}%
\ \BBA {} {Parentani}%
}{%
{Niemeyer}%
\ \BBA {} {Parentani}%
}{%
{\protect \APACyear {2001}}%
}]{%
NiemeyerRenaud:2001}
\APACinsertmetastar {%
NiemeyerRenaud:2001}%
\begin{APACrefauthors}%
{Niemeyer}, J\BPBI C.%
\BCBT {}\ \BBA {} {Parentani}, R.%
\end{APACrefauthors}%
\unskip\
\newblock
\APACrefYearMonthDay{2001}{{\APACmonth{11}}}{},
\newblock
\unskip
\newblock
\APACjournalVolNumPages{\prd}{64}{10}{101301}.
\newblock
\begin{APACrefDOI} \doi{10.1103/PhysRevD.64.101301} \end{APACrefDOI}
\PrintBackRefs{\CurrentBib}

\bibitem [\protect \citeauthoryear {%
{Noble}%
\ \BBA {} {Perelstein}%
}{%
{Noble}%
\ \BBA {} {Perelstein}%
}{%
{\protect \APACyear {2008}}%
}]{%
Noble:2008}
\APACinsertmetastar {%
Noble:2008}%
\begin{APACrefauthors}%
{Noble}, A.%
\BCBT {}\ \BBA {} {Perelstein}, M.%
\end{APACrefauthors}%
\unskip\
\newblock
\APACrefYearMonthDay{2008}{{\APACmonth{09}}}{},
\newblock
\unskip
\newblock
\APACjournalVolNumPages{\prd}{78}{6}{063518}.
\newblock
\begin{APACrefDOI} \doi{10.1103/PhysRevD.78.063518} \end{APACrefDOI}
\PrintBackRefs{\CurrentBib}

\bibitem [\protect \citeauthoryear {%
{Planck Collaboration}%
\ \protect \BOthers {.}}{%
{Planck Collaboration}%
\ \protect \BOthers {.}}{%
{\protect \APACyear {2018}}%
}]{%
Planck:2018}
\APACinsertmetastar {%
Planck:2018}%
\begin{APACrefauthors}%
{Planck Collaboration}%
, {Aghanim}, N.%
, {Akrami}, Y.%
\ et al.\end{APACrefauthors}%
\unskip\
\newblock
\APACrefYearMonthDay{2018}{{\APACmonth{07}}}{},
\newblock
\unskip
\newblock
\APACjournalVolNumPages{arXiv e-prints}{}{}{arXiv:1807.06209}.
\PrintBackRefs{\CurrentBib}

\bibitem [\protect \citeauthoryear {%
{Weinberg}%
}{%
{Weinberg}%
}{%
{\protect \APACyear {1972}}%
}]{%
Weinberg:1972}
\APACinsertmetastar {%
Weinberg:1972}%
\begin{APACrefauthors}%
{Weinberg}, S.%
\end{APACrefauthors}%
\unskip\
\newblock
\APACrefYear{1972},
\newblock
\APACrefbtitle {{Gravitation and Cosmology: Principles and Applications of the
  General Theory of Relativity}} {{Gravitation and Cosmology: Principles and
  Applications of the General Theory of Relativity}}.
\PrintBackRefs{\CurrentBib}

\bibitem [\protect \citeauthoryear {%
{Weir}%
}{%
{Weir}%
}{%
{\protect \APACyear {2018}}%
}]{%
Weir:2018}
\APACinsertmetastar {%
Weir:2018}%
\begin{APACrefauthors}%
{Weir}, D\BPBI J.%
\end{APACrefauthors}%
\unskip\
\newblock
\APACrefYearMonthDay{2018}{{\APACmonth{01}}}{},
\newblock
\unskip
\newblock
\APACjournalVolNumPages{Philosophical Transactions of the Royal Society of
  London Series A}{376}{2114}{20170126}.
\newblock
\begin{APACrefDOI} \doi{10.1098/rsta.2017.0126} \end{APACrefDOI}
\PrintBackRefs{\CurrentBib}

\bibitem [\protect \citeauthoryear {%
{Zhu}%
, {Wang}%
, {Kirsten}%
, {Cleaver}%
\BCBL {}\ \BBA {} {Sheng}%
}{%
{Zhu}%
\ \protect \BOthers {.}}{%
{\protect \APACyear {2016}}%
}]{%
Zhu:2016}
\APACinsertmetastar {%
Zhu:2016}%
\begin{APACrefauthors}%
{Zhu}, T.%
, {Wang}, A.%
, {Kirsten}, K.%
, {Cleaver}, G.%
\BCBL {}\ \BBA {} {Sheng}, Q.%
\end{APACrefauthors}%
\unskip\
\newblock
\APACrefYearMonthDay{2016}{{\APACmonth{06}}}{},
\newblock
\unskip
\newblock
\APACjournalVolNumPages{\prd}{93}{12}{123525}.
\newblock
\begin{APACrefDOI} \doi{10.1103/PhysRevD.93.123525} \end{APACrefDOI}
\PrintBackRefs{\CurrentBib}

\end{thebibliography}

\end{document}